\documentclass{elsart5p}
\usepackage{graphics}
\usepackage{graphicx}
\usepackage{amssymb}
\begin{document}

\begin{frontmatter}

\title{Dissipation effects in percolating quantum Ising magnets}

\author{Jos\'e A. Hoyos} and
\author{Thomas Vojta\corauthref{vojta}}
\ead{vojtat@umr.edu}

\corauth[vojta]{Corresponding author. Tel: (573) 341-4793 fax: (573) 341-4715}

\address{Department of Physics, University of Missouri-Rolla, Rolla, MO 65409, USA.}

\begin{abstract}
We study the effects of dissipation on a randomly dilute transverse-field
Ising magnet at and close to the percolation threshold. For weak transverse
fields, a novel percolation quantum phase transition separates a super-paramagnetic
cluster phase from an inhomogeneously ordered ferromagnetic phase.
The properties of this transition are dominated by large frozen and
slowly fluctuating percolation clusters. Implementing numerically
a strong-disorder real space renormalization group technique, we compute
the low-energy density of states which is found to be in good agreement with the analytical prediction.
\end{abstract}
\end{frontmatter}

In dilute quantum magnets, the combination of geometric and quantum
fluctuations can lead to unconventional low-temperature properties
such as singular thermodynamic quantities in quantum Griffiths phases
as well as exotic scaling at the quantum phase transitions~\cite{vojta06}.
In many real systems, the magnetic degrees of freedom are coupled
to an environment of ``heat bath'' modes. The dissipation caused by
the bath is known to qualitatively change the properties even of a
single quantum spin~\cite{leggett}. Here, we show that even small dissipation dramatically changes phases and phase transitions of a randomly dilute quantum Ising magnet.

We consider the $d$-dimensional site-dilute random transverse-field
Ising model given by the Hamiltonian
\begin{equation}
H_{I}=-\sum_{\left\langle i,j\right\rangle }J_{ij}\kappa_{i}\kappa_{j}\sigma_{i}^{z}\sigma_{j}^{z}-\sum_{i}h_{i}\kappa_{i}\sigma_{i}^{x},\label{eq:H}
\end{equation}
where the Pauli matrices $\sigma_{i}^{z}$ and $\sigma_{i}^{x}$ represent
the spin components at site $i$, the exchange interaction $J_{ij}>0$
couples nearest neighbor sites, and the transverse field $h_{i}$
controls the quantum fluctuations. Dilution is introduced via random
variables $\kappa_{i}$ which can take the values 0 and 1 with probabilities
$p$ and $1-p$, respectively.
We now couple each spin to a local ohmic bath of harmonic oscillators,

\begin{equation}
H=H_{I}+\sum_{i,n}\left[\nu_{i,n}a_{i,n}^{\dagger}a_{i,n}+\frac{1}{2}\lambda_{i,n}\sigma_{i}^{z}(a_{i,n}^{\dagger}+a_{i,n})\right],\label{eq:Hamiltonian}
\end{equation}
where $a_{i,n}$ and $a_{i,n}^{\dagger}$ are the annihilation and
creation operators of the $n$-th oscillator coupled to spin $i$;
$\nu_{i,n}$ is its frequency, and $\lambda_{i,n}$ is the coupling
constant. The baths have spectral functions ${\cal E}_{i}(\omega)=\pi\sum_{n}\lambda_{i,n}^{2}\delta(\omega-\nu_{i,n})/\nu_{i,n}=2\pi\alpha_{i}\omega e^{-\omega/\Omega_{i}}$
with $\alpha_{i}$ the dimensionless dissipation strength and $\Omega_{i}$
the cutoff energy.

For small transverse fields in the absence of dissipation ($\alpha_{i}=0$),
this system undergoes an unconventional phase transition controlled
by an infinite-randomness fixed-point ($p=p_{c}$) from a ferromagnetic
phase ($p<p_{c}$) to a quantum Griffiths paramagnetic phase ($p>p_{c}$),
where $p_{c}$ is the geometric percolation threshold.
Close to the transition~\cite{senthil-sachdev}, the low-energy density
of states (DOS) $\rho\left(\epsilon\right)$ is dominated by large
clusters tunnelling with an exponentially
small frequency, $\Delta_{s}\approx h\exp\left\{ -Bs\right\} $, where
$s$ is the cluster size, $B=\ln\left(J/h\right)$,
and $J$ and $h$ are the typical values of $J_{ij}$ and $h_{i}$,
respectively. Averaging over the cluster size distribution $n_{s}$~\cite{stauffer}
yields (in the paramagnetic phase)
\begin{equation}
\rho\left(\epsilon\right)=\sum_{s}n_{s}\delta\left(\epsilon-\Delta_{s}\right)\sim\frac{1}{\epsilon^{1-d/z^{'}}},
\label{eq:rho-a0-p>pc}
\end{equation}
where $z^{'}\sim\xi^{D}$ is the dynamical exponent, $\xi$ is the geometric
correlation length and $D$ is the fractal dimension of the infinite
cluster. $n_{s}\left(t\right)=s^{-\tau_{{\rm c}}}f\left(ts^{\sigma_{{\rm c}}}\right)$,
where $t=p-p_{c}$, $\tau_{{\rm c}}$ and $\sigma_{{\rm c}}$ are
classical percolation exponents and $f$ is a universal function: $f(x)\sim \{ -(c_{1}x)^{1/\sigma_{{\rm c}}} \} $, for $x>0$ and
$f(x)\sim \exp \{ -(c_{2}x)^{(1-1/d)/\sigma_{{\rm c}}} \} $,
for $x<0$, with $c_{1}$ and $c_{2}$ being nonuniversal constants
of order unity~\cite{stauffer}. At the percolation threshold, $z$ is formally
infinity. In this case, because geometrical fluctuations are maximal,
\begin{equation}
\rho\left(\epsilon\right)\sim\frac{1}{\epsilon\ln^{\tau_{{\rm c}}}(h/\epsilon)}
\label{eq:rho-a0-ppc}
\end{equation}
becomes even more singular. Below $p_c$, the infinite cluster
contributes with a delta peak at zero energy to the  DOS.

Dissipation dramatically changes the above scenario (for instance, consider the case
where $\alpha_{i}=\alpha$ and $\Omega_{i}=\Omega$). It was shown that for small fields
the effective dissipation strength is additive~\cite{hoyos-vojta-a}, i.e.,
the effective dissipation strength of a cluster of size $s$ is $\alpha_{s}=\alpha s$.
Therefore, a cluster of size $s$ tunnels with a renormalized frequency
$\tilde{\Delta}_{s}\sim\Delta_{s}(\Delta_{s}/\Omega)^{\alpha_{s}/\left(1-\alpha_{s}\right)}=h\exp\{ -bs/(1-\alpha s)\} $,
if $\alpha_{s}<1$. If $\alpha_{s}>1$, the cluster becomes static,
i.e., $\tilde{\Delta}_{s}=0$~\cite{leggett}. The constant $b=B+\alpha\ln(\Omega/h)$.
Note that dissipation strongly suppresses the tunnelling of large clusters of sizes near $s_{c}=1/\alpha$. This
qualitatively changes the  physical behavior in comparison
to the dissipationless case. The paramagnetic quantum Griffiths phase
is replaced at low-energies by a classical super-paramagnet phase
in which large clusters behave nearly classically~\cite{hoyos-vojta-a}.

Computing the low-energy DOS, we find that
\begin{equation}
\rho\left(\epsilon\right)=\sum_{s<s_{c}}n_{s}\delta\left(\epsilon-\tilde{\Delta}_{s}\right)\sim\frac{1}{\epsilon\ln^{\phi}\left(h/\epsilon\right)},
\label{eq:rho-adif0}
\end{equation}
with $\phi=2$, for energies $\epsilon<\epsilon_{{\rm cross}}$, where
$\epsilon_{{\rm cross}}=h\exp\left\{ -b/\alpha\right\} $ is a crossover
energy scale above which weak dissipation becomes unimportant and
$\rho\left(\epsilon\right)$ reduces to
Eq.~(\ref{eq:rho-a0-p>pc})
or (\ref{eq:rho-a0-ppc}). Furthermore, the singular dependence of
the DOS Eq.~(\ref{eq:rho-adif0}) does \emph{not} depend on the $p$, meaning
it is the same below, at and above to the percolation threshold.
The infinite cluster (when $p<p_{c}$) and the frozen clusters (when $s>s_{c}$)
contribute to a delta peak to the  DOS. Thermodynamical
quantities in the above scenario show unconventional behavior as discussed in Ref.~\cite{hoyos-vojta-a}.

We now use a generalization of the strong-disorder renormalization
group method in order to compute the  DOS~\cite{fisher,schehr-rieger}.
We search for the strongest bond in the system $\max \{ h_{i},\, J_{ij}\} $.
(a) In the case of a coupling, say $J_{23}$, spins $\sigma_{2}$
and $\sigma_{3}$ become locked in a ferromagnetic state, and thus,
are replaced by a single effective spin $\tilde{\sigma}_{2}$. The renormalized transverse field reads $\tilde{h}_{2}=h_{2}h_{3}(J_{2}/\Omega_{2})^{\alpha_{2}}
(J_{2}/\Omega_{3})^{\alpha_{3}}/J_{2}$.
The effective dissipation strength is $\tilde{\alpha}_{2}=\alpha_{2}+\alpha_{3}$ and the new cutoff $\tilde{\Omega}_{2}=J_{2}$. The coupling between
the new spin and spin $\sigma_{i}$ is $\tilde{J}_{2i}=J_{2i}+J_{3i}$.
(b) If the strongest bond is a field, say $h_{2}$, then we proceed
as follows. (b.i) If $\Omega_{2}>h_{2}$, we integrate out the oscillators
until $\Omega_{2}=h_{2}$. Hence, we renormalize $\tilde{h}_{2}=h_{2}(h_{2}/\Omega_{2})^{\alpha_{2}}$
and $\tilde{\Omega}_{2}=h_{2}$. (b.ii) If $\Omega_{2}=h_{2}$, we
freeze the spin $\sigma_{2}$ in the direction of the transverse field
$h_{2}$. Then, we remove it from the system and renormalize the remaining couplings $\tilde{J}_{ij}=J_{ij}+J_{2i}J_{2j}/h_{2}$.

\begin{figure}[ht]
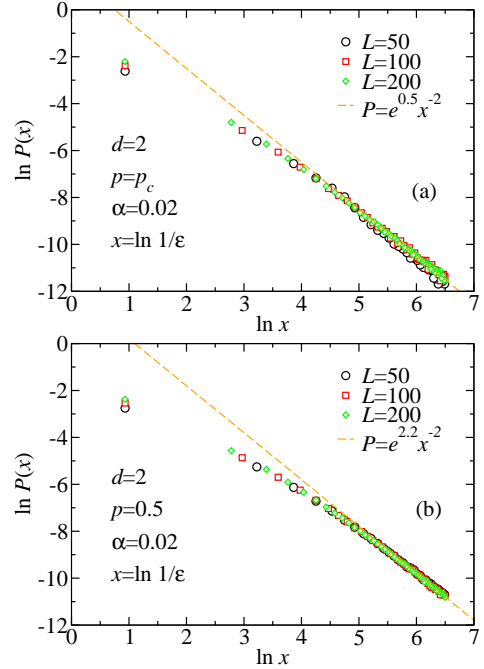

\begin{center}
\includegraphics[clip,width=0.7\columnwidth,
  keepaspectratio]{DOS-p=pc.eps}\\
\includegraphics[clip,width=0.7\columnwidth,
  keepaspectratio]{DOS-p=0.5.eps}
\end{center}

\caption{Distribution of low-energy tunneling rates (a) for $p=p_{c}=0.407253$ and (b)
$p=0.5$ for different system sizes $L$. The distribution were built decimating $2\,000$
systems for $L=200$ and 100, and $8\,000$ systems for $L=50$. \label{cap:P(x)}}
\end{figure}

Iterating these steps, the DOS is obtained by building the histogram of effective
tunneling rates $\epsilon$ of the percolation clusters. Fig.~\ref{cap:P(x)} shows the
distribution $P\left(x\right)$, where $x=\ln(1/\epsilon)$, for $p=p_{c}$ and $p=0.5$ for
a two dimensional square lattice of size $L$x$L$, and coupling constants $J_{ij}$
(transverse fields $h_{i}$) uniformly distributed between 0 and 1 (0 and 0.1).
According to Eq.~(\ref{eq:rho-adif0}), $P(x)\sim x^{-\phi}$, with $\phi=2$ regardless the
value of $p$. As shown by the fits in Fig.~\ref{cap:P(x)}, the data are in very good
agreement with this analytical prediction.
Moreover, for $p=0.5\neq p_c$, the data disagree with the prediction of the undamped
system Eq.~(\ref{eq:rho-a0-p>pc}), showing the importance of the dissipation for the
low-energy DOS.

In conclusion, we have shown that dilute quantum Ising magnets with Ohmic dissipation
undergo a novel percolation quantum phase transition between a ferromagnetic and a
classical super-paramagnetic phase. The low-energy DOS displays striking singularities
due to the suppression of tunneling by damping. Using a strong-disorder renormalization
group method, we have computed the low-energy DOS numerically and found it to be in good
agreement with the analytical prediction.

This work was supported by NSF grant no. DMR-0339147 and by Research
Corporation.

\end{document}